\documentclass[twoside]{article}
\usepackage{Proc_NTSE}

 \usepackage{graphicx,amsmath,amssymb}
\usepackage{indentfirst}
 \usepackage[usenames]{color}
 \usepackage[colorlinks=true, urlcolor=navyblue, linkcolor=navyblue, citecolor=navyblue]{hyperref}
\usepackage{epstopdf}
\usepackage{appendix}

\definecolor{navyblue}{rgb}{0,0.08,0.45}

\def\Dslash{\raise.15ex\hbox{/}\kern-.7em D}
\def\Pslash{\raise.15ex\hbox{/}\kern-.7em P}

\newcommand{\beq}{\begin{equation}}
\newcommand{\enq}{\end{equation}}
\newcommand{\beqa}{\begin{eqnarray}}
\newcommand{\beqast}{\begin{eqnarray*}}
\newcommand{\enqa}{\end{eqnarray}}
\newcommand{\enqast}{\end{eqnarray*}}
\newcommand{\beml}{\begin{multline}}
\newcommand{\enml}{\end{multline}}

\newcommand{\bec}{\begin{center}}
\newcommand{\enc}{\end{center}}
\newcommand{\beqo}{\begin{quote}}
\newcommand{\enqo}{\end{quote}}

\newcommand{\half}{{\textstyle{\frac{1}{2}}}}

\newcommand{\mbf}[1]{\mathbf{#1}}

\begin{document}
\thispagestyle{plain}

\begin{flushright}
{
SLAC-PUB-15735       ~~ ~~~~\\
\date{today}}
\end{flushright}

\begin{center}
{\Large \bf \strut
Light-Front  Holographic  Quantum Chromodynamics
\strut}\\
\vspace{10mm}
{\large \bf 
Stanley J. Brodsky$^{a}$, Guy F. de T\'eramond$^{b}$, and  Hans G\"unter Dosch{$^c$}}
\end{center}

\noindent%

\begin{center}
\small $^a${\it SLAC National Accelerator Laboratory, Stanford University, Stanford, CA 94309, USA} \\
{\small $^b$ \it Universidad de Costa Rica, San Jos\'e, Costa Rica} \\
{\small $^c$\it Institut f\"ur Theoretische Physik, Philosophenweg 16, D-6900 Heidelberg, Germany} \\ 
\end{center}
\markboth{
S. J. Brodsky,  G. de T\'eramond, and H. G. Dosch}
{
Light-Front Holographic QCD} 

\begin{abstract}
Light-Front Hamiltonian theory, derived from the quantization of the QCD Lagrangian at fixed light-front time $\tau = x^0+x^3$, provides a rigorous frame-independent framework for solving nonperturbative QCD.  The eigenvalues of the light-front QCD Hamiltonian $H_{LF} $ predict the hadronic mass spectrum, and the corresponding eigensolutions provide the light-front wavefunctions which describe hadron structure.  In the case of mesons, the valence Fock-state wavefunctions of $H_{LF}$ for zero quark mass satisfy a single-variable relativistic equation of motion in the invariant variable $\zeta^2=b^2_\perp x(1-x)$, which is conjugate to the invariant mass squared ${M^2_{q \bar q} }$. The effective confining potential $U(\zeta^2)$  in this frame-independent ``light-front Schr\"odinger equation" systematically incorporates the effects of higher quark and gluon Fock states.  Remarkably,  the potential has a unique form of a harmonic oscillator potential if one requires that the chiral QCD  action remains conformally invariant. The result is a nonperturbative relativistic light-front quantum mechanical wave equation which incorporates color confinement and other essential spectroscopic and dynamical features of hadron physics. 

Anti-de Sitter space in five dimensions plays a special role in elementary particle physics since it provides an exact geometrical representation of the conformal group.  Remarkably, gravity  in AdS$_5$ space is holographically dual to frame-independent light-front Hamiltonian theory.  Light-front holography also leads to a precise relation between the bound-state amplitudes in the fifth dimension $z$ of AdS space and the variable $\zeta$, the argument of the boost-invariant light-front wavefunctions describing the internal structure of hadrons in physical space-time.    The holographic mapping of gravity in AdS space to QCD with a specific ``soft-wall" dilaton yields the confining potential $U(\zeta^2)$ which is consistent with conformal invariance of the QCD action and the light-front Schr\"odinger equation, extended to hadrons with arbitrary spin $J$.  One thus obtains an effective light-front effective theory for general spin which respects the conformal symmetry of the four-dimensional classical QCD Lagrangian. The predictions of the LF equations of motion include a zero-mass pion in the chiral $m_q\to 0$ limit, and linear Regge trajectories ${M}^2(n,L) \propto n+L$  with the same slope  in the radial quantum number $n$ and the orbital angular momentum $L$. 
The light-front AdS/QCD holographic approach thus gives a frame-independent representation of color-confining dynamics, Regge spectroscopy, as well as the excitation spectra of relativistic light-quark meson and also baryon bound states in QCD in terms of a single mass parameter.  

We also briefly discuss the implications of the underlying conformal template of QCD  for renormalization scale-setting and the implications of light-front quantization for the value of the cosmological constant. 
\\[\baselineskip] 
\end{abstract}

\section{Introduction}

The remarkable advantages  of using light-front time $\tau = x^0+x^3/c$ (the ``front form") to quantize a theory instead of the standard time $t=x^0$ (the ``instant form") was first demonstrated by Dirac.   As Dirac showed~\cite{Dirac:1949cp}, the front form has the maximum number of kinematic generators of the Lorentz group, including the boost operator.  Thus the description of a hadron at fixed $\tau$ is independent of the observer's Lorentz frame, making it ideal for addressing dynamical processes in quantum chromodynamics.

\begin{figure}[h]
\centering
\includegraphics[width=14.00cm]{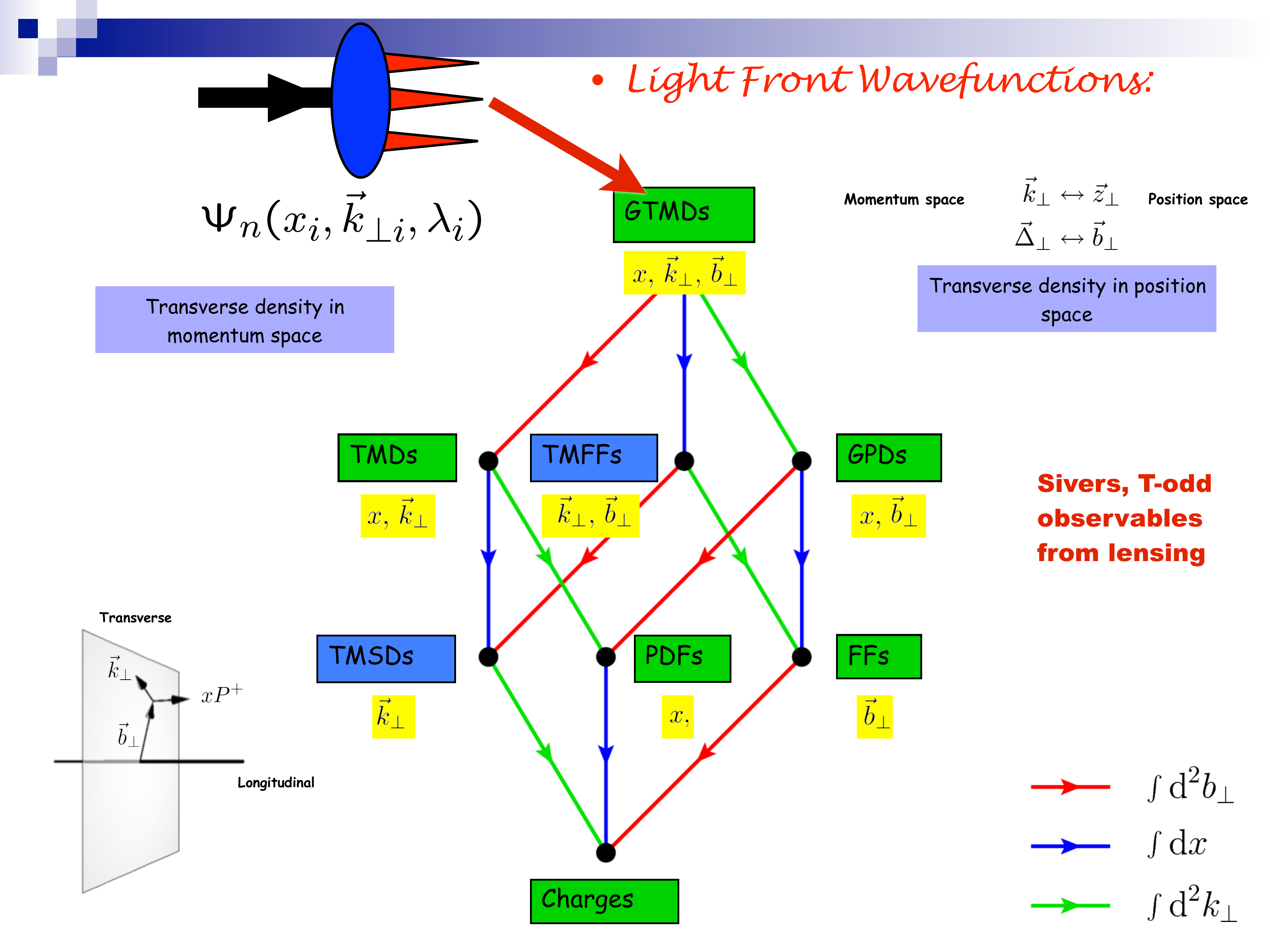}\hspace{0pt}
\caption{\small Examples of hadronic observables, including Wigner functions  and $T$-odd observables which are based on overlaps of light-front wavefunctions. 
Adopted from a figure by F. Lorce and B.~Pasquini.~\cite{Lorce:2012jy}}
\label{Lorce}
\end{figure}

The quantization of QCD at fixed light-front (LF)  time -- light-front quantization -- provides a first-principles method for solving nonperturbative QCD.  Given the Lagrangian, one can determine the LF Hamiltonian $H_{LF}$ in terms of the independent quark and gluon fields. The eigenvalues of  $H_{LF}$  determine the mass-squared values of both the discrete and continuum hadronic spectra. The eigensolutions determine the LF wavefunctions required for predicting hadronic phenomenology. 
The LF method is relativistic, has no fermion-doubling,  is formulated in Minkowski space, and is frame-independent.  
The eigenstates  are defined at fixed $\tau$  within the causal horizon, so that causality is maintained without normal-ordering.   In fact, light-front physics is a fully relativistic field theory, but its structure is similar to nonrelativistic atomic physics, and the resulting bound-state equations can be formulated as relativistic Schr\"odinger-like equations at equal light-front time. 
Given the frame-independent light-front wavefunctions (LFWFs) $\psi_{n/H}$ , one can compute a large range of hadronic
observables, starting with form factors, structure functions, generalized parton distributions, Wigner distributions, etc., as illustrated in  Fig. \ref{Lorce}. 
For example, the ``handbag" contribution~\cite{Brodsky:2000xy} to the  $E$ and $H$ generalized parton distributions for deeply virtual Compton scattering can be computed from the overlap of LFWFs, automatically satisfy the known sum rules.

Computing hadronic matrix elements of currents is particularly simple in the light-front, since they can be written  as an overlap of light-front wave functions (LFWFs) as in the Drell-Yan-West 
 formula~\cite{Drell:1969km,West:1970av,Brodsky:1980zm}. 
For example,  a virtual photon couples only to forward-moving $k^+ > 0$ quarks, and only processes with the same number of initial and final partons are allowed.  In contrast,  if one uses ordinary fixed time $t$,  the hadronic states must be boosted  from the hadron's rest frame to a moving frame -- an intractable dynamical problem. 
In fact, the boost of a composite system at fixed time $t$ is only known at weak binding~\cite{Brodsky:1968ea,Brodsky:1968xc}.  Moreover, form factors at fixed instant time $t$  require computing off-diagonal matrix elements as well as the contributions of currents arising from fluctuations of the vacuum in the initial state which connect to the hadron wavefunction in the final state. Thus, the knowledge of wave functions alone is not sufficient to compute covariant current matrix elements in the usual instant-form framework.  

The gauge-invariant meson and baryon distribution amplitudes which control hard exclusive and direct reactions are the valence LFWFs integrated over transverse momentum at fixed $x_i= {k^+ / P^+}$.   The  ERBL evolution of distribution amplitudes and the factorization theorems for hard exclusive processes  were derived using LF theory~\cite{Lepage:1980fj,Efremov:1979qk}.

Because of Wick's theorem, light-front time-ordered perturbation theory is equivalent to covariant Feynman perturbation theory.   The higher order calculation of the electron anomalous moment at order $\alpha^3$ and the ``alternating denominator method" for renormalizing LF perturbation theory is given in Ref. ~\cite{Brodsky:1973kb}. 

Quantization in the light-front provides the rigorous field-theoretical realization of the intuitive ideas of the parton model~\cite{Feynman:1969ej, Feynman:1973xc} which is formulated at fixed $t$ in the infinite-momentum frame~\cite{Fubini:1965xx, Weinberg:1966jm}.  The same results are obtained in the front form for any frame; e.g., the structure functions and other probabilistic parton distributions measured in deep inelastic scattering are obtained from the squares of the boost invariant
LFWFs, the eigensolution of the light-front Hamiltonian.  The familiar kinematic variable $x_{bj}$ of deep inelastic scattering becomes identified with the LF fraction at small $x.$ 

A measurement in the front form is analogous to taking a flash picture. The image in the resulting photograph records the state of the object as the front of  the light wave from the flash illuminates it; in effect, this is  a measurement  within the spacelike causal horizon ${\Delta x_\mu}^2 \le 0.$  Similarly, measurements such as deep inelastic lepton-proton scattering $\ell H \to \ell^\prime X$, determine  the LFWF and structure structures of the  target hadron $H$ at fixed light-front time. 
For example, the BFKL Regge behavior of structure functions can be demonstrated~\cite{Mueller:1993rr} from the behavior of LFWFs at small $x$.

One can also prove fundamental theorems for relativistic quantum field theories using the front form, including: (a) the cluster decomposition theorem~\cite{Brodsky:1985gs} and (b) the vanishing of the anomalous gravitomagnetic moment for any Fock state of a hadron~\cite{Brodsky:2000ii}; one also can show that a nonzero anomalous magnetic moment of a bound state  requires nonzero angular momentum of the constituents. 
Stasto and Cruz-Santiago~\cite{Cruz-Santiago:2013vta} have shown that the cluster properties~\cite{Antonuccio:1997tw} of LF time-ordered perturbation theory, together with $J^z$ conservation, can be used to elegantly derive the Parke-Taylor rules for multi-gluon scattering amplitudes.
 The counting-rule~\cite{Brodsky:1994kg} behavior of structure functions at large $x$ and Bloom-Gilman duality have also been derived in LFQCD as well as from holographic QCD~\cite{Gutsche:2013zia}.

The physics of diffractive deep inelastic scattering and other hard processes where the projectile hadron remains intact is most easily analysed using LF QCD~\cite{Brodsky:2002ue}.     The existence of ``lensing effects"  at leading twist, such as the $T$-odd ``Sivers effect" in spin-dependent semi-inclusive deep-inelastic scattering,  was first demonstrated using LF methods~\cite{Brodsky:2002cx}.   QCD properties such as  ``color transparency"~\cite{Brodsky:1988xz}, the ``hidden color" of the deuteron LFWF~\cite{Brodsky:1983vf}, and the existence of intrinsic  heavy quarks in the LFWFs of light hadrons~\cite{Brodsky:1980pb,Franz:2000ee} can be derived from the structure of hadronic LFWFs.
It is also possible to compute jet hadronization at the amplitude level from first principles from the LFWFs~\cite{Brodsky:2008tk}. The LFWFs of hadrons thus provide a direct connection between observables and the QCD Lagrangian.

Light-front quantization is thus the natural framework for the description  the nonperturbative relativistic bound-state structure of hadrons in  quantum chromodynamics.   The formalism is rigorous, relativistic and frame-independent. 
In principle, one can solve nonperturbative QCD by diagonalizing the light-front QCD Hamiltonian $H_{LF}$ directly using the ``discretized light-cone quantization" (DLCQ) method~\cite{Brodsky:1997de} which imposes periodic boundary conditions to discretize the $k^+$ and $k_\perp$ momenta, or the Hamiltonian transverse lattice formulation
introduced in refs.~\cite{Bardeen:1976tm,Burkardt:2001mf,Bratt:2004wq}. The hadronic spectra and light-front  wavefunctions are then obtained from the eigenvalues and eigenfunctions of the Heisenberg problem $H_{LF} \vert \psi \rangle = M^2 \vert \psi \rangle$, an infinite set of coupled integral equations for the light-front components $\psi_n = \langle n \vert \psi \rangle$ in a Fock expansion~\cite{Brodsky:1997de}. These nonperturbative methods have the advantage that they are frame-independent,  defined in physical Minkowski space-time, and have no fermion-doubling problem.  
The DLCQ method has been applied successfully in lower space-time dimensions~\cite{Brodsky:1997de}, such as QCD(1+1)~\cite{Pauli:1985pv}. 
It has also been applied successfully to a range of 1+1 string theory problems by Hellerman and Polchinski~\cite{Hellerman:1999nr,Polchinski:1999br}.

Solving the eigenvalue problem using DLCQ is a formidable computational task  for a non-abelian quantum field theory  in four-dimensional space-time because of the large number of independent variables. Consequently, alternative methods and approximations  are necessary to better understand the nature of relativistic bound-states in the strong-coupling regime. 
One of the most promising methods for solving nonperturbative (3+1)  QCD is the ``Basis Light-Front Quantization" (BFLQ) method initiated by James Vary~\cite{Vary:2009gt} and his collaborators.  In the BLFQ method one constructs a complete orthonormal basis of eigenstates  based on the eigensolutions of the  effective light-front Schr\"odinger equation derived from light-front holography, in the spirit of the nuclear shell model. Matrix diagonalization for BLFQ should converge more rapidly than DLCQ since the basis states have a mass spectrum  close to the observed hadronic spectrum.   

An extensive review of light-front quantization is given in Ref.~\cite{Brodsky:1997de}  
As we shall discuss here, light-front quantized field theory in physical $3+1$ space-time has a holographic dual with dynamics of theories in five-dimensional anti-deSitter space, giving important insight into the nature of color confinement in QCD.

\section{What is the origin of the QCD mass scale?}

If one sets the masses of the quarks to zero, no mass scale appears explicitly in the QCD Lagrangian.  The classical theory thus displays invariance under both scale  (dilatation) and special conformal transformations~\cite{Parisi:1972zy}.   Nevertheless, the quantum theory built upon this conformal template displays color confinement,  a mass gap, as well as asymptotic freedom. A  fundamental question is thus how does the mass scale which determines the masses of the light-quark hadrons,  the range of color confinement, and the running of the  coupling appears in QCD?

A hint to the origin of the mass scale in nominally conformal theories was given in 1976 in a remarkable paper by  V.~de Alfaro, S.~Fubini and G.~Furlan (dAFF)~\cite{deAlfaro:1976je} in the context of one-dimensional 
quantum mechanics.  They showed that the mass scale which 
breaks dilatation invariance can appear in the equations of motion without violating  the conformal invariance of the action.  In fact, this is only possible if the resulting potential has the form of a confining harmonic oscillator, and the transformed time variable $\tau$ that appears in the confining theory has a limited range.  

In this contribution to the NTSE meeting we will review how the application of the dAFF procedure,
together with light-front quantum mechanics and light-front holographic mapping,
leads to a new analytic approximation to QCD -- a light-front Hamiltonian and corresponding one-dimensional light-front (LF) Schr\"odinger and Dirac equations which are frame-independent, relativistic, and reproduce crucial features of the spectroscopy and dynamics of the light-quark hadrons.
The predictions of the LF equations of motion include a zero-mass pion in the chiral $m_q\to 0$ limit, and linear Regge trajectories ${M}^2(n,L) \propto n+L$  with the same slope  in the
radial quantum number $n$ (the number of nodes) and 
$L= {\rm max} \,|L^z|$, the internal orbital angular momentum.    In fact, we will also show that the effective  confinement potential which appears in the LF equations of motion is unique if we require that the chiral QCD action remains conformally invariant.

\section{Light-Front Holography}

An important  analysis tool for QCD is Anti-de Sitter space in five dimensions.  In particular, AdS$_5$ provides a remarkable geometric representation of the conformal group which 
underlies the conformal symmetry of classical QCD. One can modify AdS space by using a dilaton factor in the AdS
action 
$e^{\varphi(z)} $ to introduce the QCD 
confinement scale.  However, we shall show that
if one imposes the requirement that  the action of the corresponding one-dimensional effective theory  remains conformal invariant,
then the dilaton profile  $\varphi(z) \propto z^s$ is constrained to have the specific power $s = 2$,  a remarkable result which follows from the dAFF construction of conformally invariant quantum mechanics~\cite{Brodsky:2013kpr}.   A related argument is given in 
Ref.~\cite{Glazek:2013jba}  The quadratic form 
$\varphi(z) = \pm \, \kappa^2 z^2$
immediately leads to linear Regge trajectories~\cite{Karch:2006pv} in the hadron mass squared.   

A simple way to obtain confinement and discrete normalizable modes is to truncate AdS space with the introduction of a sharp cut-off in the infrared region of AdS space, as in the ``hard-wall'' model~\cite{Polchinski:2001tt},  where one considers  a slice  of AdS space, $0 \leq z \leq z_0$, and imposes boundary conditions on the fields at the IR border $z_0 \sim 1/\Lambda_{\rm QCD}$.  As first shown by Polchinski and Strassler~\cite{Polchinski:2001tt}, the modified AdS space, provides a derivation of dimensional counting rules~\cite{Brodsky:1973kr, Matveev:1973ra} in QCD for the leading power-law fall-off of hard scattering beyond the perturbative regime. The modified theory generates the point-like hard behavior expected from QCD, instead of the soft behavior characteristic of extended objects~\cite{Polchinski:2001tt}.   The physical  states in  AdS space are represented by normalizable modes $\Phi_P(x,z) = e^{-iP \cdot x} \Phi(z)$, with plane waves along Minkowski coordinates $x^\mu$ and a profile function $\Phi(z)$ along the holographic coordinate $z$. The hadronic invariant mass $P_\mu P^\mu = M^2$  is found by solving the eigenvalue problem for the AdS wave equation.

``Light-Front Holography" refers to the remarkable fact that dynamics in AdS space in five dimensions is dual to 
a semiclassical approximation to
Hamiltonian theory in physical  $3+1$ space-time quantized at fixed light-front time~\cite{deTeramond:2008ht}.  The  correspondence between AdS and QCD, which was originally motivated by
the AdS/CFT correspondence between gravity on a higher-dimensional 
space and conformal field theories 
in physical space-time~\cite{Maldacena:1997re},  has its most explicit and simplest realization as a direct holographic mapping to light-front Hamiltonian theory~\cite{deTeramond:2008ht}. 
For example, the equation of motion for mesons on the light-front has exactly the same single-variable form as the AdS equation of motion; one can then interpret the AdS fifth dimension variable $z$ in terms of the physical variable $\zeta$, representing the invariant separation of the $q$ and $\bar q$ at fixed light-front time.  There is a precise 
connection between the quantities that enter the fifth 
dimensional AdS space and the physical variables of LF theory.  The AdS mass  parameter $\mu R$ maps to the LF orbital angular momentum.  The formulae for electromagnetic~\cite{Polchinski:2002jw} and gravitational~\cite{Abidin:2008ku} form factors in AdS space map to the exact Drell-Yan-West formulae in light-front QCD~\cite{Brodsky:2006uqa, Brodsky:2007hb, Brodsky:2008pf}.  

\begin{figure}[h]
\centering
\includegraphics[width=6.00cm]{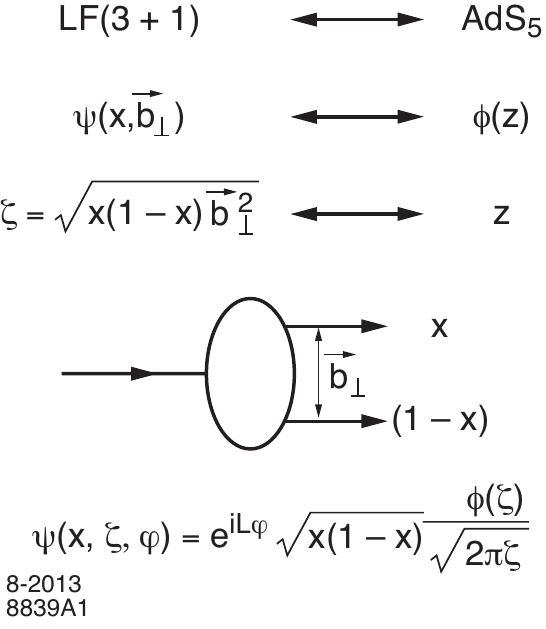}\hspace{0pt}
\caption{\small Light-Front Holography: Mapping between the hadronic wavefunctions of the Anti-de Sitter approach and eigensolutions of the light-front Hamiltonian theory  derived from the equality of LF and AdS  formula for EM and gravitational current matrix elements and their identical equations of motion.}
\label{dictionary}
\end{figure}

The light-front holographic principle provides a precise relation between the bound-state amplitudes in AdS space and the boost-invariant LF wavefunctions describing the internal structure of hadrons in physical space-time  (See Fig. \ref {dictionary}). The resulting valence Fock-state wavefunctions
satisfy a single-variable relativistic equation of motion analogous to the eigensolutions of the nonrelativistic radial Schr\"odinger equation. 
The quadratic dependence in the effective quark-antiquark potential 
$U(\zeta^2,J) =  \kappa^4 \zeta^2 +2 \kappa^2(J-1)$ 
is determined uniquely from conformal invariance.   The  constant term $ 2 \kappa^2(J-1) = 2 \kappa^2(S+L-1)  $  is fixed by the duality between AdS and LF quantization for spin-$J$ states, a correspondence which follows specifically from the separation of kinematics and dynamics on the light-front~\cite{deTeramond:2013it}.  
The LF potential thus has a specific power dependence--in effect, it is a light-front harmonic oscillator potential.
It  is confining and reproduces the observed linear Regge behavior of the light-quark hadron spectrum in both the orbital angular momentum $L$ and the radial node number $n$. The  pion is predicted to be massless in the chiral limit ~\cite{deTeramond:2009xk}  - the positive contributions to $m^2_\pi$ from the LF potential and kinetic energy 
is cancelled by the constant term  in $U(\zeta^2,J)$ for $J=0.$   This holds for the positive sign of the dilaton profile  $\varphi(z) =  \kappa^2 z^2$.
The  LF dynamics retains conformal invariance of the action despite the presence of a fundamental mass scale.  
The constant term in the  LF potential $U(\zeta^2,J)$ derived from LF Holography is essential;  the masslessness of the pion and the separate dependence on $J$ and $L$ are  consequences of the potential derived from the holographic LF duality with AdS for general  $J$ and $L$~\cite{Brodsky:2013kpr, deTeramond:2013it}. Thus the light-front holographic approach provides an analytic frame-independent first approximation to the color-confining dynamics,  spectroscopy, and excitation spectra of the relativistic light-quark bound states of QCD.   It is systematically improvable in full QCD using  the basis light-front quantization (BLFQ) method~\cite{Vary:2009gt} and other methods. 

We now give an example of  light-front holographic mapping for the specific case of the elastic pion form factor.
In the higher-dimensional gravity theory, the hadronic transition amplitude  corresponds to
the  coupling of an external electromagnetic field $A^M(x,z)$,  for a photon propagating in AdS space, with an extended field $\Phi_P(x,z)$ describing a meson in AdS is~\cite{Polchinski:2002jw}
 \begin{equation} \label{MFF}
 \int \! d^4x \, dz  \sqrt{g} \, A^M(x,z)
 \Phi^*_{P'}(x,z) \overleftrightarrow\partial_M \Phi_P(x,z) \\
  \sim
 (2 \pi)^4 \delta^4 \left( P'  \! - P - q\right) \epsilon_\mu  (P + P')^\mu F_M(q^2) ,
 \end{equation}
where the coordinates of AdS$_5$ are the Minkowski coordinates $x^\mu$ and $z$ labeled $x^M = (x^\mu, z)$,
 with $M, N = 1, \cdots 5$,  and $g$ is the determinant of the metric tensor. 
The expression on the right-hand side
of (\ref{MFF}) represents the space-like QCD electromagnetic transition amplitude in physical space-time
$
\langle P' \vert J^\mu(0) \vert P \rangle = \left(P + P' \right)^\mu F_M(q^2).
$
It is the EM matrix element of the quark current  $J^\mu = \sum_q e_q \bar q \gamma^\mu q$, and represents a local coupling to pointlike constituents. Although the expressions for the transition amplitudes look very different, one can show  that a precise mapping of the matrix elements  can be carried out at fixed light-front time~\cite{Brodsky:2006uqa, Brodsky:2007hb}.

The form factor is computed in the light front formalism from the matrix elements of the plus current $J^+$
 in order to avoid coupling to Fock states with different numbers of constituents and is given by the Drell-Yan-West expression. The form factor can  be conveniently written in impact space as a sum of overlap of LFWFs of the $j = 1,2, \cdots, n-1$ spectator constituents~\cite{Soper:1976jc} 
\begin{equation} \label{eq:FFb}
F_M(q^2) =  \sum_n  \prod_{j=1}^{n-1}\int d x_j d^2 \mbf{b}_{\perp j}
\exp \! {\Bigl(i \mbf{q}_\perp \! \cdot \sum_{j=1}^{n-1} x_j \mbf{b}_{\perp j}\Bigr)}
 \\ \times
\left\vert  \psi_{n/M}(x_j, \mbf{b}_{\perp j})\right\vert^2,
\end{equation}
corresponding to a change of transverse momentum $x_j \mbf{q}_\perp$ for each
of the $n-1$ spectators with  $\sum_{i = 1}^n \mbf{b}_{\perp i} = 0$.  The formula is exact if the sum is over all Fock states $n$.

For simplicity, consider a two-parton bound-state. The $q \bar q$ LF Fock state wavefunction for a meson  can be written as
\begin{equation} \label{psi}
\psi(x,\zeta, \varphi) = e^{i L \varphi} X(x) \frac{\phi(\zeta)}{\sqrt{2 \pi \zeta}},
\end{equation}
thus factoring the longitudinal, $X(x)$,  transverse  $\phi(\zeta)$ and angular dependence $\varphi$.
 If both expressions for the form factor are to be
identical for arbitrary values of $Q$, we obtain $\phi(\zeta) = (\zeta/R)^{3/2} \Phi(\zeta)$ and $X(x) = \sqrt{x(1-x)}$~\cite{Brodsky:2006uqa},
where we identify the transverse impact LF variable $\zeta$ with the holographic variable $z$,
$z \to \zeta = \sqrt{x(1-x)} \vert \mbf b_\perp \vert$, where $x$ is the longitudinal momentum fraction and $ b_\perp$ is  the transverse-impact distance between the quark and antiquark. Extension of the results to arbitrary $n$ follows from the $x$-weighted definition of the transverse impact variable of the $n-1$ spectator system given in Ref.  \cite{Brodsky:2006uqa}.  Identical results follow from mapping the matrix elements of the energy-momentum tensor~\cite{Brodsky:2008pf}.

\section{The Light-Front Schr\"odinger Equation: A Semiclassical Approximation to QCD \label{LFQCD}}

It is advantageous to reduce the full multiparticle eigenvalue problem of the LF Hamiltonian to an effective light-front Schr\"odinger equation  which acts on the valence sector LF wavefunction and determines each eigensolution separately~\cite{Pauli:1998tf}.   In contrast,  diagonalizing the LF Hamiltonian yields all eigensolutions simultaneously, a complex task.
The central problem 
then becomes the derivation of the effective interaction $U$ which acts only on the valence sector of the theory and has, by definition, the same eigenvalue spectrum as the initial Hamiltonian problem.  In order to carry out this program one must systematically express the higher Fock components as functionals of the lower ones. This  method has the advantage that the Fock space is not truncated, and the symmetries of the Lagrangian are preserved~\cite{Pauli:1998tf}.

A hadron has four-momentum $P = (P^-, P^+,  \mbf{P}_\perp)$, $P^\pm = P^0 \pm P^3$ and invariant mass $P^2 = M^2$. The generators $P = (P^-, P^+,  \vec{P}_\perp)$ are constructed canonically from the QCD Lagrangian by quantizing the system on the light-front at fixed LF time $x^+$, $x^\pm = x^0 \pm x^3$~\cite{Brodsky:1997de}. 
The LF Hamiltonian $P^-$ generates the LF time evolution with respect to the LF time  $x^+$, whereas the LF longitudinal $P^+$ and transverse momentum $\vec P_\perp$ are kinematical generators.

\begin{figure}[h]
\centering
\includegraphics[width=12cm]{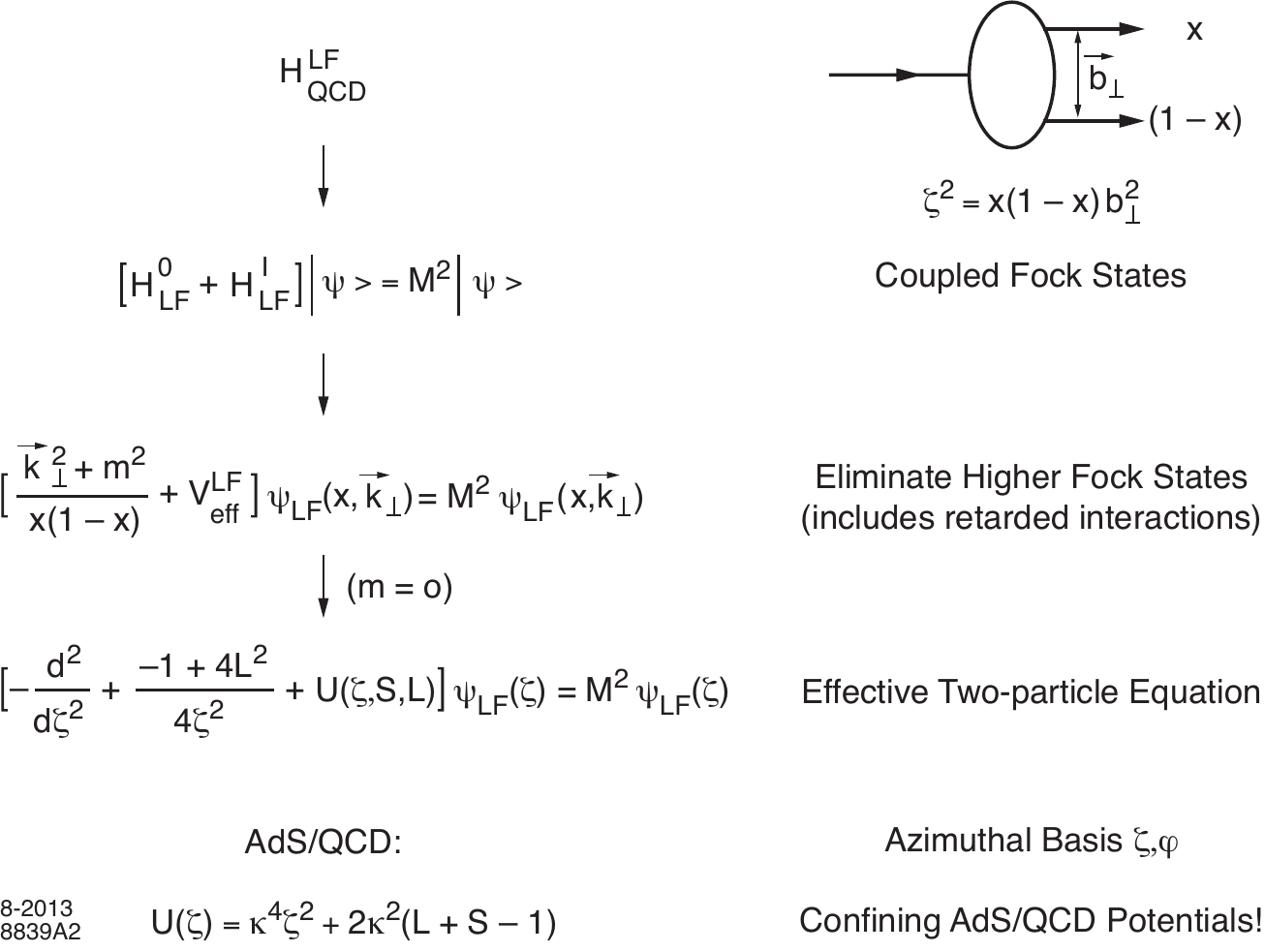}\hspace{0pt}
 \caption{\small Reduction of the QCD light-front Hamiltonian to an effective $q \bar q$ bound state equation. The potential is determined from spin-$J$ representations on AdS$_5$ space. The harmonic oscillator form of $U(\zeta^2)$ is determined by the requirement that the action remain conformally invariant.}
\label{reduction}
\end{figure}

In the limit of zero quark masses the longitudinal modes decouple  from the  invariant  LF Hamiltonian  equation  $H_{LF} \vert \phi \rangle  =  M^2 \vert \phi \rangle$,  with  $H_{LF} = P_\mu P^\mu  =  P^- P^+ -  \mbf{P}_\perp^2$.  The result is a relativistic and frame-independent light-front  wave equation for $\phi$~\cite{deTeramond:2008ht} (See Fig. \ref{reduction})
\begin{equation} \label{LFWE}
\left[-\frac{d^2}{d\zeta^2}
- \frac{1 - 4L^2}{4\zeta^2} + U\left(\zeta^2, J\right) \right]
\phi_{n,J,L}(\zeta^2) = 
M^2 \phi_{n,J,L}(\zeta^2).
\end{equation}
 This equation describes the spectrum of mesons as a function of $n$, the number of nodes in $\zeta$, the total angular momentum  $J$, which represent the maximum value of $\vert J^z \vert$, $J = \max \vert J^z \vert$,
and the internal orbital angular momentum of the constituents $L= \max \vert L^z\vert$.
The variable $z$ of AdS space is identified with the LF   boost-invariant transverse-impact variable $\zeta$~\cite{Brodsky:2006uqa}, 
thus giving the holographic variable a precise definition in LF QCD~\cite{deTeramond:2008ht, Brodsky:2006uqa}.
For a two-parton bound state $\zeta^2 = x(1-x) b^{\,2}_\perp$.
In the exact QCD theory $U$ is related to the two-particle irreducible $q \bar q$ Green's function.

The potential in the the Light-Front Schr\"odinger equation (\ref{LFWE}) is determined from the two-particle irreducible (2PI) $ q \bar q \to q \bar q $ Greens' function.  In particular, the reduction from higher Fock states in the intermediate states
leads to an effective interaction $U\left(\zeta^2, J\right)$  for the valence $\vert q \bar q \rangle$ Fock state~\cite{Pauli:1998tf}.
A related approach for determining the valence light-front wavefunction and studying the effects of higher Fock states without truncation has been given in Ref.~\cite{Chabysheva:2011ed}.

Unlike ordinary instant-time quantization, the light-front Hamiltonian equations of motion are frame independent; remarkably, they  have a structure which matches exactly the eigenmode equations in AdS space. This makes a direct connection of QCD with AdS methods possible.  In fact, one can
derive the light-front holographic duality of AdS  by starting from the light-front Hamiltonian equations of motion for a relativistic bound-state system
in physical space-time~\cite{deTeramond:2008ht}.

\section{Effective Confinement from the Gauge/Gravity Correspondence}

Recently we have derived wave equations for hadrons with arbitrary spin $J$ starting from an effective action in  AdS space~\cite{deTeramond:2013it}.    An essential element is the mapping of the higher-dimensional equations  to the LF Hamiltonian equation  found in Ref.~\cite {deTeramond:2008ht}.  This procedure allows a clear distinction between the kinematical and dynamical aspects of the LF holographic approach to hadron physics.  Accordingly, the non-trivial geometry of pure AdS space encodes the kinematics,  and the additional deformations of AdS encode the dynamics, including confinement~\cite{deTeramond:2013it}.

A spin-$J$ field in AdS$_{d+1}$ is represented by a rank $J$ tensor field $\Phi_{M_1 \cdots M_J}$, which is totally symmetric in all its indices.  In presence of a dilaton background field $\varphi(z)$ the effective action is~\cite{deTeramond:2013it} 
\begin{equation}
\label{Seff}
S_{\it eff} = \int d^{d} x \,dz \,\sqrt{\vert g \vert}  \; e^{\varphi(z)} \,g^{N_1 N_1'} \cdots  g^{N_J N_J'} 
  \Big(  g^{M M'} D_M \Phi^*_{N_1 \dots N_J}\, D_{M'} \Phi_{N_1 ' \dots N_J'}  
 - \mu_{\it eff}^2(z)  \, \Phi^*_{N_1 \dots N_J} \, \Phi_{N_1 ' \dots N_J'} \Big),
 \end{equation}
where 
$D_M$ is the covariant derivative which includes parallel transport. 
The effective mass  $\mu_{\it eff}(z)$, which encodes kinematical aspects of the problem, is an {\it a priori} unknown function,  but the additional symmetry breaking due to its $z$-dependence allows a clear separation of kinematical and dynamical effects~\cite{deTeramond:2013it}.
The dilaton background field $\varphi(z)$ in  (\ref{Seff})   introduces an energy scale in the five-dimensional AdS action, thus breaking conformal invariance. It  vanishes in the conformal ultraviolet limit $z \to 0$.

 A physical hadron has plane-wave solutions and polarization indices along the 3 + 1 physical coordinates
 $\Phi_P(x,z)_{\nu_1 \cdots \nu_J} = e^{ i P \cdot x} \Phi_J(z) \epsilon_{\nu_1 \cdots \nu_J}({P})$,
 with four-momentum $P_\mu$ and  invariant hadronic mass  $P_\mu P^\mu \! = M^2$. All other components vanish identically. 
 The wave equations for hadronic modes follow from the Euler-Lagrange equation for tensors orthogonal to the holographic coordinate $z$,  $\Phi_{z N_2 \cdots N_J}  = 0$. Terms in the action which are linear in tensor fields, with one or more indices along the holographic direction, $\Phi_{z N_2 \cdots N_J}$, give us 
 the kinematical constraints required to eliminate the lower-spin states~\cite{deTeramond:2013it}.  Upon variation with respect to $ \hat \Phi^*_{\nu_1 \dots \nu_J}$,
 we find the equation of motion~\cite{deTeramond:2013it}  
\begin{equation}  \label{PhiJM}
 \left[ 
   -  \frac{ z^{d-1- 2J}}{e^{\varphi(z)}}   \partial_z \left(\frac{e^{\varphi(z)}}{z^{d-1-2J}} \partial_z   \right) \\
  +  \frac{(m\,R )^2}{z^2}  \right]  \Phi_J = M^2 \Phi_J,
  \end{equation}
  with  $(m \, R)^2 =(\mu_{\it eff}(z) R)^2  - J z \, \varphi'(z) + J(d - J +1)$,
  which is  the result found in Refs.~\cite{deTeramond:2008ht, deTeramond:2012rt} by rescaling the wave equation for a scalar field.  Similar results were found
  in Ref.~\cite{Gutsche:2011vb}. Upon variation with respect to
$ \hat \Phi^*_{N_1 \cdots z  \cdots N_J}$  we find the kinematical constraints which  eliminate lower spin states from the symmetric field tensor~\cite{deTeramond:2013it}  
\begin{equation} \label{sub-spin}
 \eta^{\mu \nu } P_\mu \,\epsilon_{\nu \nu_2 \cdots \nu_J}({P})=0, \quad
\eta^{\mu \nu } \,\epsilon_{\mu \nu \nu_3  \cdots \nu_J}({P})=0.
 \end{equation}

Upon the substitution of the holographic variable $z$ by the LF invariant variable $\zeta$ and replacing
  $\Phi_J(z)   = \left(R/z\right)^{J- (d-1)/2} e^{-\varphi(z)/2} \, \phi_J(z)$ 
in (\ref{PhiJM}), we find for $d=4$ the LF wave equation (\ref{LFWE}) with effective potential~\cite{deTeramond:2010ge}
\begin{equation} \label{U}
U(\zeta^2, J) = \frac{1}{2}\varphi''(\zeta^2) +\frac{1}{4} \varphi'(\zeta^2)^2  + \frac{2J - 3}{2 \zeta} \varphi'(\zeta^2) ,
\end{equation}
provided that the AdS mass $m$ in (\ref{PhiJM}) is related to the internal orbital angular momentum $L = max \vert L^z \vert$ and the total angular momentum $J^z = L^z + S^z$ according to $(m \, R)^2 = - (2-J)^2 + L^2$.  The critical value  $L=0$  corresponds to the lowest possible stable solution, the ground state of the LF Hamiltonian.
For $J = 0$ the five dimensional mass $m$
 is related to the orbital  momentum of the hadronic bound state by
 $(m \, R)^2 = - 4 + L^2$ and thus  $(m\, R)^2 \ge - 4$. The quantum mechanical stability condition $L^2 \ge 0$ is thus equivalent to the Breitenlohner-Freedman stability bound in AdS~\cite{Breitenlohner:1982jf}.

\begin{figure}[h]
\centering
\includegraphics[width=6.80cm]{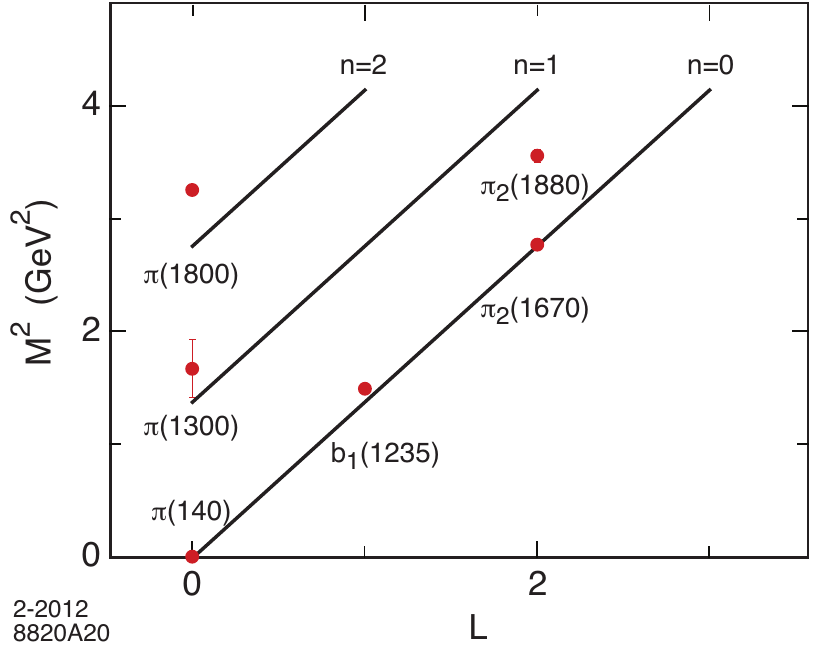} \hspace{20pt}
\includegraphics[width=6.80cm]{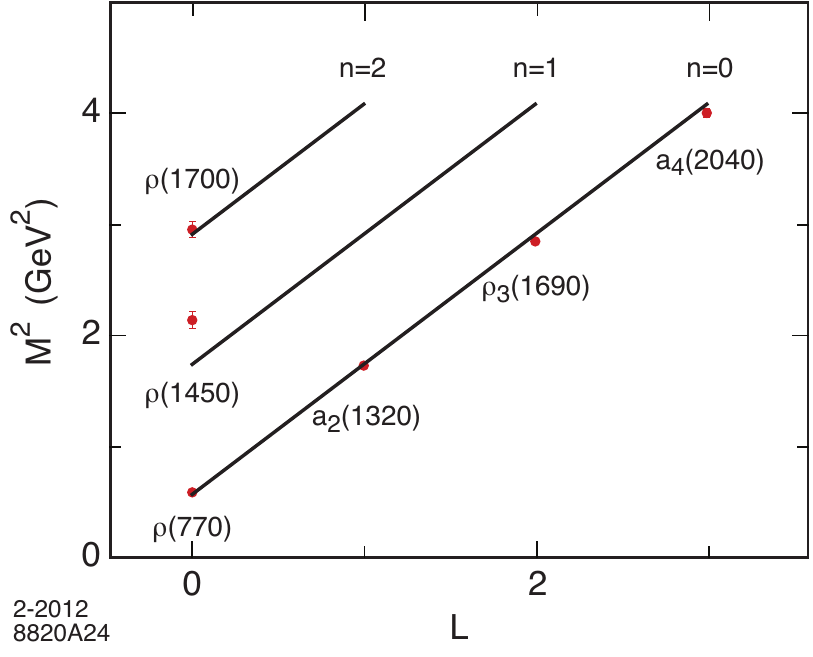}
 \caption{\small $I = 1$ parent and daughter Regge trajectories for the $\pi$-meson family (left) with
$\kappa= 0.59$ GeV; and  the   $\rho$-meson
 family (right) with $\kappa= 0.54$ GeV.}
\label{pionspec}
\end{figure} 

\begin{figure}[h]
\centering
\includegraphics[width=6.80cm]{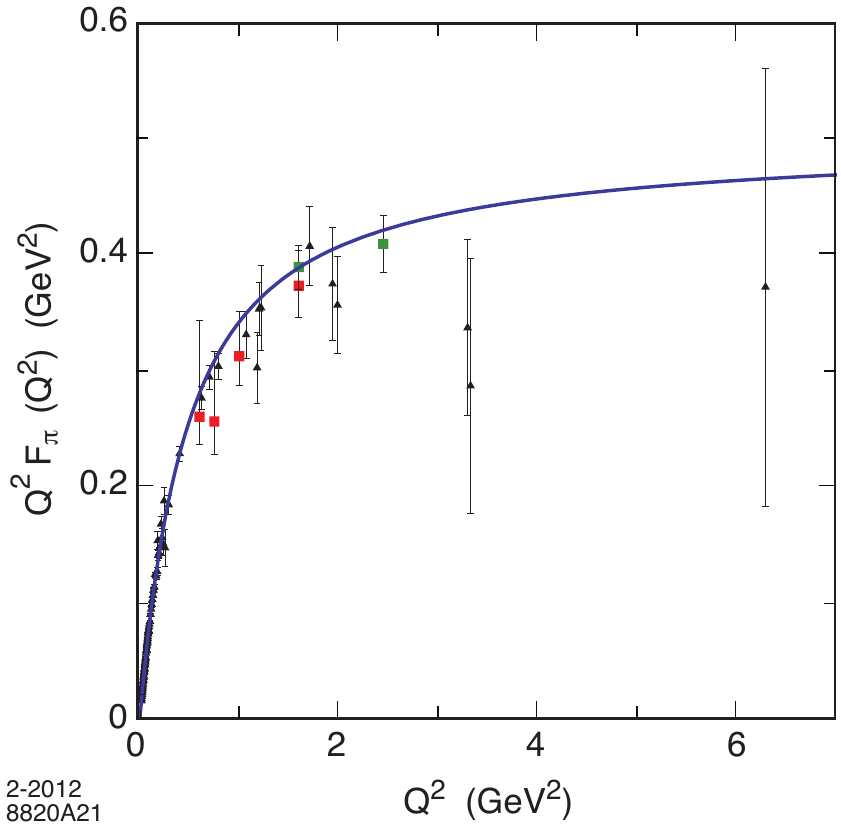} \hspace{0pt}
 \caption{\small Light-front holographic prediction for the space-like pion form factor.  }
\label{pionff}
\end{figure} 

The effective interaction $U(\zeta^2,J)$
is instantaneous in LF time and acts on the lowest state of the LF Hamiltonian.  This equation describes the spectrum of mesons as a function of $n$, the number of nodes in $\zeta^2$,
 the internal orbital angular momentum $L = L^z$, and the total angular momentum $J=J^z$,
with $J^z = L^z + S^z$  the sum of the  orbital angular momentum of the constituents and their internal spins.
The  ${\rm SO(2)}$ Casimir  $L^2$  corresponds to  the group of rotations in the transverse LF plane.
The LF wave equation is the relativistic frame-independent front-form analog of the non-relativistic radial Schr\"odinger equation for muonium  and other hydrogenic atoms in presence of an instantaneous Coulomb potential. The LF harmonic oscillator potential could in fact emerge from the exact QCD formulation when one includes contributions from the 
effective potential $U$ which are due to the exchange of two connected gluons; {\it i.e.}, ``H'' diagrams~\cite{Appelquist:1977tw}.
We notice that $U$ becomes complex for an excited state since a denominator can vanish; this gives a complex eigenvalue and the decay width.  The multi gluon exchange diagrams also could be connected to the Isgur-Paton flux-tube model of confinement; the collision of flux tubes could give rise to the ridge phenomena recently observed in high energy $pp$ collisions at RHIC~\cite{Bjorken:2013boa}.

The correspondence between the LF and AdS equations  thus determines the effective confining interaction $U$ in terms of the infrared behavior of AdS space and gives the holographic variable $z$ a kinematical interpretation. The identification of the orbital angular momentum 
is also a key element of our description of the internal structure of hadrons using holographic principles.

The dilaton profile $\exp{\left(\pm \kappa^2 z^2\right)}$  
leads to linear Regge trajectories~\cite{Karch:2006pv}.
For the  confining solution $\varphi = \exp{\left(\kappa^2 z^2\right)}$ the effective potential is
$U(\zeta^2,J) =   \kappa^4 \zeta^2 + 2 \kappa^2(J - 1)$  leads to eigenvalues
$M_{n, J, L}^2 = 4 \kappa^2 \left(n + \frac{J+L}{2} \right)$,
with a string Regge form $M^2 \sim n + L$.  
A detailed discussion of the light meson and baryon spectrum,  as well as  the elastic and transition form factors of the light hadrons using LF holographic methods, is given in 
Ref.~\cite{deTeramond:2012rt}.  As an example the spectral predictions  for the $J = L + S$ light pseudoscalar and vector meson  states are  compared with experimental data in Fig. \ref{pionspec} for the positive sign dilaton model.

\begin{figure}[h]
\centering
\includegraphics[width=6.80cm]{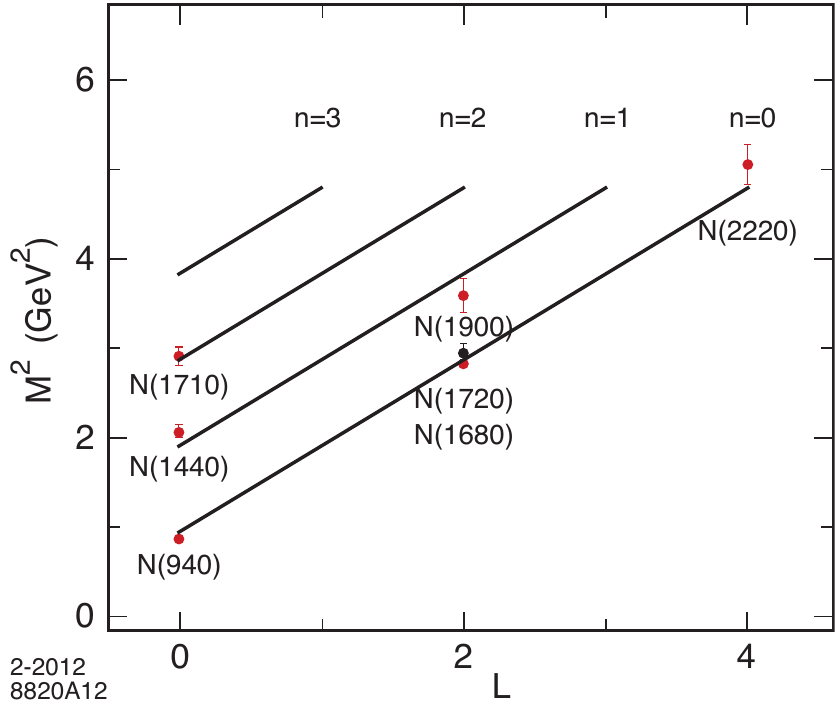} \hspace{-2pt}
\includegraphics[width=6.80cm]{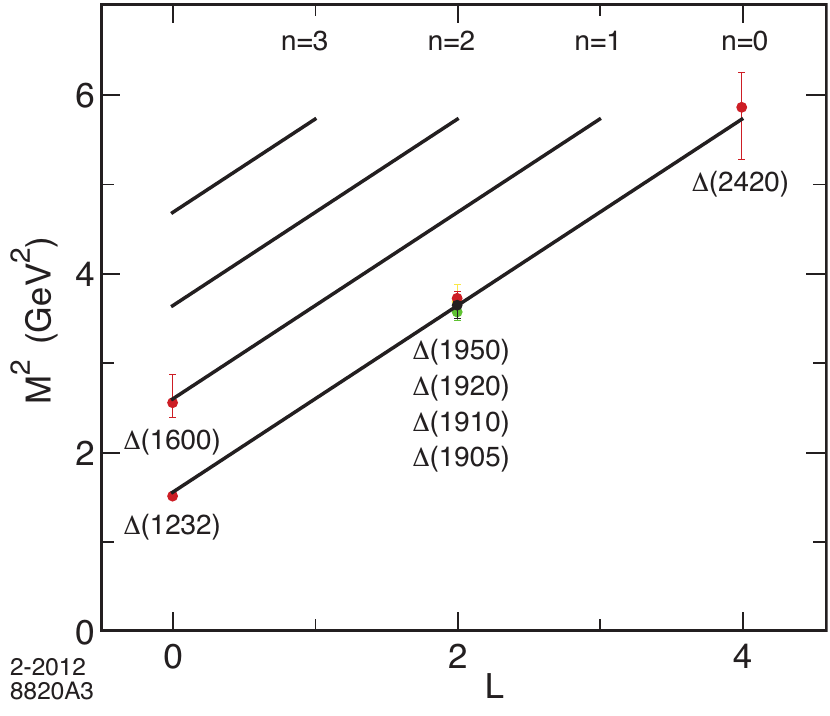} 
 \caption{\small Light front holographic predictions of the light-front Dirac equation for the nucleon spectrum. Orbital and radial excitations for the positive-parity sector are shown
 for the $N$ (left) and $\Delta$ (right) for $\kappa= 0.49$ GeV and $\kappa = 0.51$ GeV respectively. All confirmed positive and negative-parity resonances from PDG 2012 are well accounted using the procedure described in   \cite{deTeramond:2012rt}.}
\label{nucleonspect}
\end{figure} 

\begin{figure}[h]
\centering
\includegraphics[width=7.0cm]{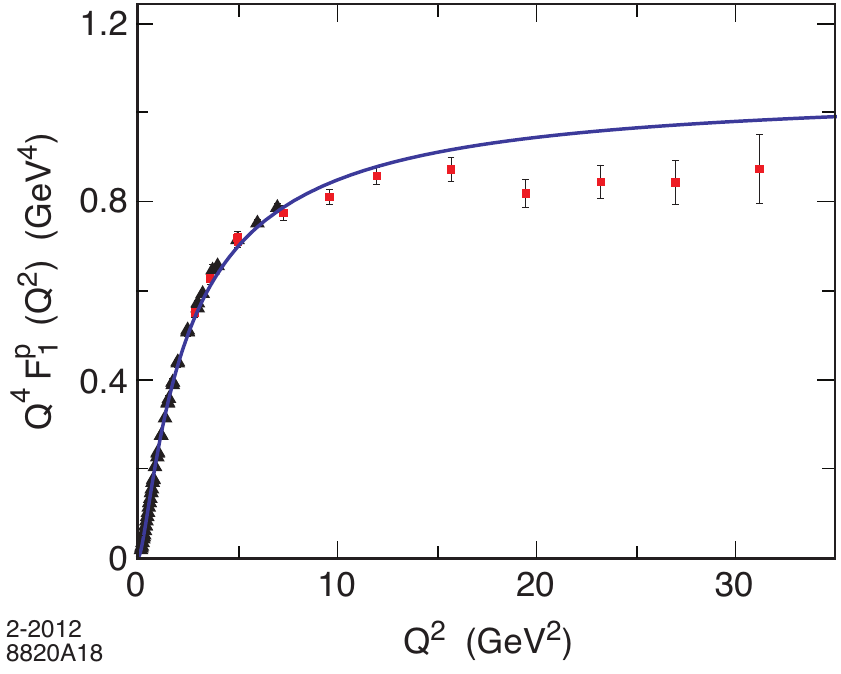}
 \hspace{2pt}
\includegraphics[width=7.0cm]{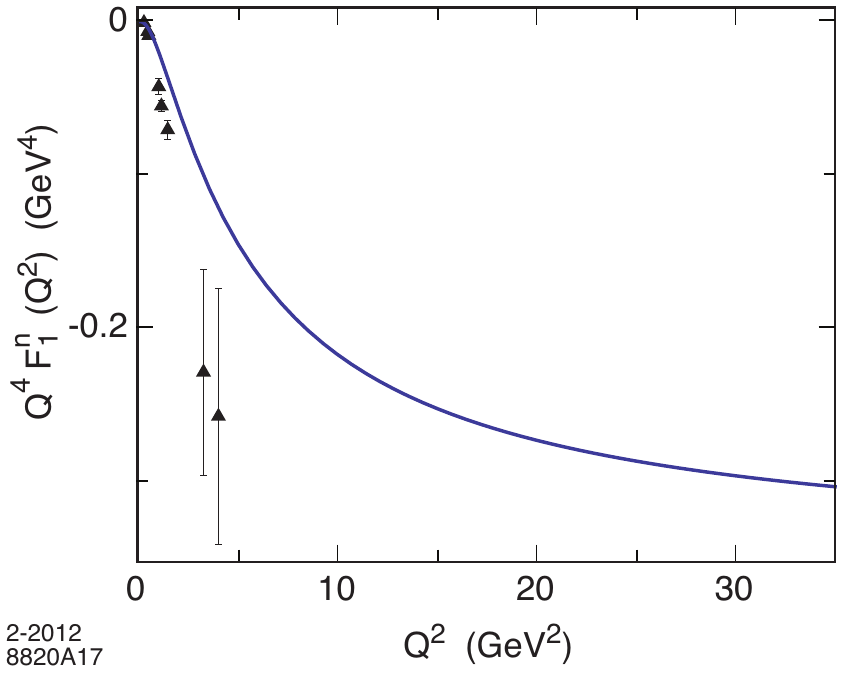}
\includegraphics[width=7.0cm]{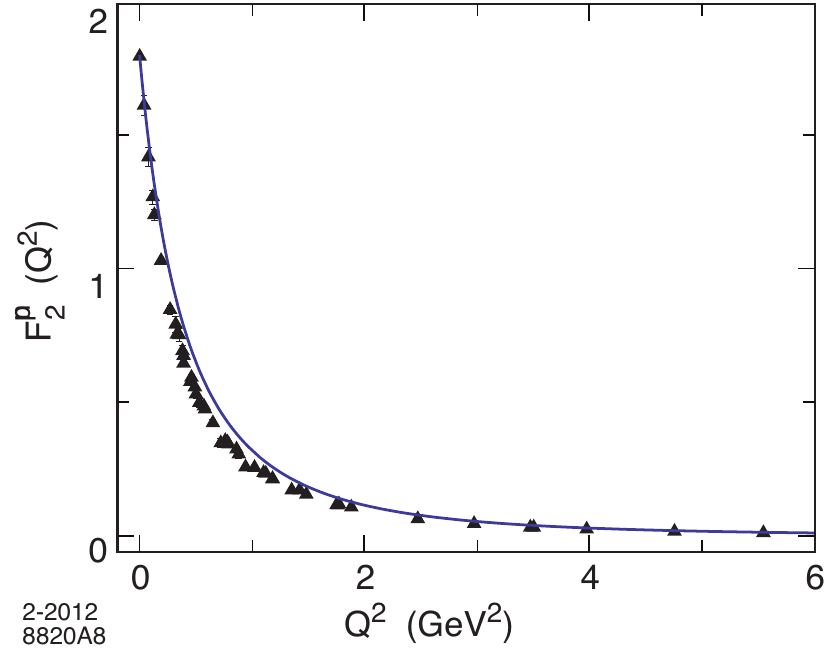} 
\hspace{2pt}
\includegraphics[width=7.0cm]{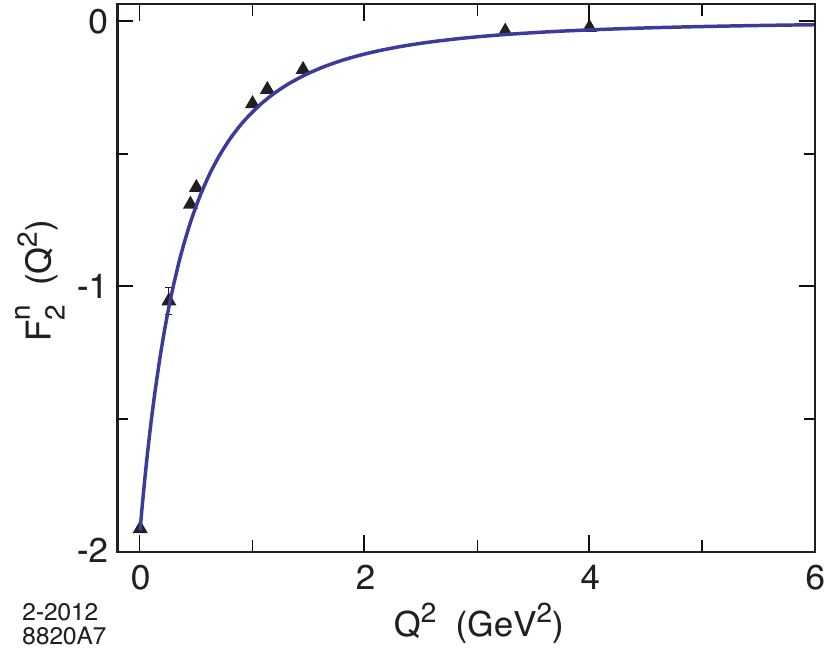} 
\caption{Light-front holographic predictions for the nucleon form factors normalized to their static values.  }
\label{nucleonffs}
\end{figure} 

The predictions of the resulting LF Schr\"odinger and Dirac equations for hadron light-quark spectroscopy and form factors for $m_q=0$ and $\kappa \simeq 0.5$ GeV are shown in Figs. \ref{pionspec}-\ref{nucleonffs} for a dilaton profile $\varphi(z) = \kappa^2 z^2$. A detailed discussion of the computations is given in Ref.  \cite{deTeramond:2012rt}.

\section{Uniqueness of the Confining Potential}

If one starts with a dilaton profile  $e^{\varphi(z)}$ with $\varphi \propto z^s,$  the existence of a massless pion in the limit of massless quarks determines
uniquely the value  $s = 2$.
To show this, one can use the stationarity of bound-state energies with respect to variation of parameters.
More generally, the effective theory should incorporate the fundamental conformal symmetry of the four-dimensional  classical QCD Lagrangian in the limit of massless quarks. To this end we study the invariance properties of a one-dimensional field theory under the full conformal group following the  dAFF construction of Hamiltonian operators described in Ref. ~\cite{deAlfaro:1976je}.

One starts with the one-dimensional action
$
{ \cal{S}}= \half \int dt (\dot Q^2 - g/Q^2),
$
which  is invariant under conformal transformations in the 
variable $t$. In addition to the Hamiltonian  $H_t$ there are two more invariants of motion for this field theory, namely the 
dilation operator $D$ and $K$, corresponding to the special conformal transformations in $t$.  
Specifically, if one introduces the  the new variable $\tau$ defined through 
$d\tau= d t/(u+v\,t + w\,t^2)$ and the  rescaled fields $q(\tau) = Q(t)/(u + v\, t + w \,t^2)^{1/2}$,
it then follows that the the operator
$G= u\,H_t + v\,D + w\,K$  generates
evolution in  $\tau$~\cite{deAlfaro:1976je}.
The Hamiltonian corresponding to the operator $G$ which introduces the mass scale
is a linear combination of the old Hamiltonian $H_t$, $D$, the generator of dilations, and
$K$, the generator of special conformal transformations.
It contains the confining potential 
$(4 u w - v^2) \zeta^2/8$, that is the confining term in (\ref{U}) for a quadratic dilaton profile and thus $\kappa^4 =  (4 u w - v^2)/8$.
The variable $\tau$ is related to the variable $t$ for the case  $u w >0, \,v=0$   by 
$
\tau =\frac{1}{\sqrt{u\,w}} \arctan\left(\sqrt{\frac{w}{u}} t\right),
$
{\it i.e.}, $\tau$ has only a limited range. The finite range of invariant LF time $\tau=x^+/P^+$ can be interpreted as a feature of the internal frame-independent LF  time difference between the confined constituents in a bound state. For example, in the collision of two mesons, it would allow one to compute the LF time difference between the two possible quark-quark collisions~\cite{Brodsky:2013kpr}.

\section{The Light-Front Vacuum}

It is conventional to define the vacuum in quantum field theory as the lowest energy eigenstate of the instant-form Hamiltonian.  Such an eigenstate is defined at a single time $t$ over all space $\vec x$.  It is thus acausal and frame-dependent. The instant-form vacuum thus must be normal-ordered in order to avoid violations of causality when computing correlators and other matrix elements. In contrast, in the front form, the vacuum state is defined as the  eigenstate of lowest invariant mass $M$.   It is defined at fixed light-front time $x^+ = x^0 + x^3$ over all $x^-= x^0 - x^3$ and $\vec x_\perp$, the extent of space that can be observed within the speed of light.   It is frame-independent and only requires information within the causal horizon.

Since all particles have positive $k^+= k^0 + k^z > 0$ and $+$ momentum is conserved in the front form, the usual vacuum bubbles are kinematically forbidden in the front form.  In fact the LF vacuum for QED, QCD, and even the Higgs Standard Model is trivial up to possible zero modes -- backgrounds with zero four-momentum. In this sense it is already normal-ordered.  In the case of the Higgs theory, the usual Higgs vacuum expectation value is replaced by a classical $k^+=0$  background zero-mode field  which is not sensed by the energy momentum tensor~\cite{Srivastava:2002mw}.  The phenomenology of the Higgs theory is unchanged.

There are thus no quark or gluon vacuum condensates in the LF vacuum-- as first noted by Casher and Susskind~\cite{Casher:1974xd}; the corresponding physics is contained within the 
LFWFs themselves~\cite{Brodsky:2009zd,Brodsky:2010xf,Chang:2013pq,Chang:2013epa,Glazek:2011vg}, thus eliminating a major contribution to the cosmological constant. In the light-front formulation of quantum field theory, phenomena such as the GMOR relation -- usually  associated with condensates in the instant form vacuum --  are properties of the the hadronic LF wavefunctions themselves.   An exact Bethe-Salpeter analysis shows that the quantity that appears in the GMOR relation is the matrix element $<0|\bar \psi \gamma_5 \psi|\pi>$  for the pion to couple locally  to the vacuum via a pseudoscalar operator  -- not a vacuum expectation value $<0|\bar\psi \psi|0>$. In the front-form  $<0|\bar \psi \gamma_5 \psi|\pi>$ involves the pion LF Fock state with parallel $q$ and $\bar q$ spin and $L^z= \pm 1$.  This pion Fock state automatically appears when the quarks are massive.

The frame-independent causal front-form vacuum is a good match to the ``void" -- the observed universe without luminous matter.    Thus it is natural in the front form to obtain zero cosmological constant from quantum field theory.   

\section{The Conformal Symmetry Template}

In the case of perturbative QCD, the running coupling  $\alpha_s(Q^2)$ becomes constant in the limit of zero $\beta$-function and zero quark mass, and conformal symmetry becomes manifest.  In fact, the renormalization scale uncertainty in 
pQCD predictions can be eliminated by using the Principle of Maximum Conformality (PMC)~\cite{Brodsky:2011ig}.
Using the PMC/BLM procedure~\cite{Brodsky:1982gc}, 
all non-conformal contributions in the perturbative expansion series are summed into the running coupling by shifting the renormalization scale in $\alpha_s$ from its initial value, and one obtains unique, scale-fixed, scheme-independent predictions at any finite order.  One can also introduce a generalization of conventional dimensional regularization, 
the ${\cal R}_\delta$ schemes which illuminates the renormalization scheme and scale ambiguities of pQCD predictions, exposes the general pattern of nonconformal terms, and allows one to systematically determine the argument of the running coupling order by order in pQCD in a form which can be readily automatized~\cite{Mojaza:2012mf, Brodsky:2013vpa}.
The resulting PMC scales and  finite-order PMC predictions are to high accuracy independent of the choice of initial renormalization scale.  For example, PMC scale-setting leads to a scheme-independent 
pQCD prediction~\cite{Brodsky:2012ik}
for the top-quark forward-backward asymmetry which is within one $\sigma$ of the Tevatron measurements. 
The PMC procedure also provides scale-fixed, scheme-independent commensurate scale relations~\cite{Brodsky:1994eh}, relations between observables which are based on the underlying conformal behavior of QCD such as the generalized Crewther relation~\cite{Brodsky:1995tb}.
The PMC satisfies  all of the principles of the renormalization group: reflectivity, symmetry, and transitivity, and it thus eliminates an unnecessary source of systematic error in pQCD predictions~\cite{Wu:2013ei}.

\section{Summary}

The triple complementary connection of  (a)  AdS space,  (b) its LF holographic dual, 
and (c) the relation to the algebra of the conformal group in one dimension, is characterized by a quadratic confinement LF potential, and 
thus a dilaton profile with the power $z^s$, with the unique power $s = 2$.  In fact,  for $s=2$ the mass of the $J=L=n=0$  pion is 
automatically zero in the chiral limit.  The separate dependence on $J$ and $L$ leads to a  mass ratio of the $\rho$ and the $a_1$ mesons which coincides with the result of the Weinberg sum rules~\cite{Weinberg:1967kj}. One predicts linear  Regge trajectories  with the same slope in the relative orbital angular momentum $L$ and the 
LF radial quantum humber $n$.  The AdS approach, however,  goes beyond the purely group theoretical considerations of dAFF, since 
features such as the masslessness of the pion and the separate dependence on $J$ and $L$ are a consequence of the potential (\ref{U}) derived from the duality with AdS for general high-spin representations.

\begin{figure}[h]
\centering
\includegraphics[width=8.00cm]{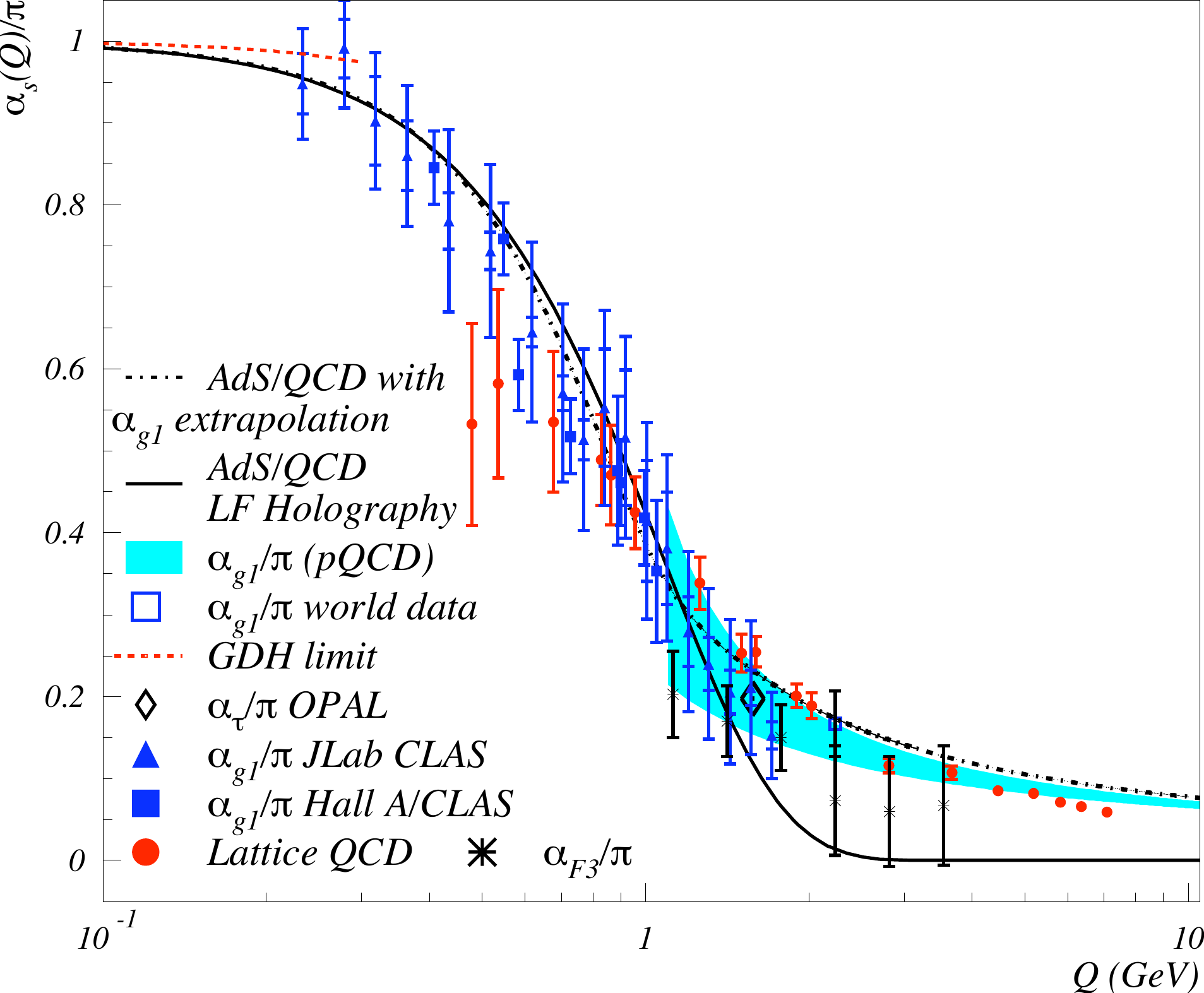} 
 \caption{\small Light-front holographic results for the QCD running coupling from Ref. ~\cite{Brodsky:2010ur} normalized to $\alpha_s(0)/ \pi =1.$  
 The result is analytic, defined at all scales and exhibits an infrared fixed point.}
 \label{alphas}
\end{figure} 

The QCD mass scale $\kappa$ in units of GeV has to be determined by one measurement; e.g., the pion decay constant $f_\pi.$  All other masses and size parameters are then predicted. The running of the QCD coupling  is predicted in the infrared region for $Q^2 <  4 \kappa^2$  to have the  form $\alpha_s(Q^2) \propto \exp{\left(-Q^2\over 4 \kappa^2\right)}$. As shown in Fig. \ref{alphas}, the result agrees with the shape of the effective charge defined from the Bjorken sum rule~\cite{Brodsky:2010ur}, displaying an infrared fixed point~\cite{Brodsky:2010ur}.  In the nonperturbative domain soft gluons are in effect sublimated into the effective confining potential. Above this region, hard-gluon exchange becomes important, leading to asymptotic freedom.  
The scheme-dependent scale $\Lambda_{QCD}$ that appears in the QCD running coupling in any given renormalization scheme  could be determined in terms of $\kappa.$

In our previous papers we have applied  LF holography to  baryon spectroscopy, space-like and time-like form factors, as well as transition amplitudes such as 
$\gamma^* \gamma \to \pi^0$, $\gamma^* N \to N^*$, all based on essentially one mass scale parameter $\kappa.$  Many other applications have been presented in the literature, including recent results by Forshaw and Sandapen~\cite{Forshaw:2012im} for diffractive  $\rho$ electroproduction,  based on the light-front holographic prediction for the longitudinal $\rho$ LFWF. Other recent applications include predictions for  generalized parton distributions (GPDs)~\cite{Vega:2010ns}, and a model for nucleon and flavor form factors~\cite{Chakrabarti:2013dda}.

The treatment of the chiral limit in the LF holographic approach to strongly coupled QCD is substantially different from the standard approach  based on chiral perturbation theory.
In the conventional approach, 
spontaneous symmetry breaking  by a  non-vanishing chiral quark condensate $\langle  \bar \psi \psi  \rangle$ plays the crucial role.  In QCD sum  rules \cite{Shifman:1978bx}  $\langle  \bar \psi \psi \rangle$ brings in non-perturbative elements into the perturbatively calculated spectral sum rules.  It should  be noted, however, that the definition of the condensate, even in lattice QCD necessitates a renormalization procedure for the operator product, and it is not a directly observable quantity.  In contrast, in  Bethe-Salpeter~\cite{Maris:1997hd} and light-front analyses~\cite{Brodsky:2012ku}, the Gell Mann-Oakes-Renner relation~\cite{GellMann:1968rz}
for $m^2_\pi/m_q$ involves the decay matrix element $\langle 0 |\bar \psi\gamma_5 \psi |\pi \rangle$ instead of $\langle 0| \bar \psi \psi | 0\rangle$.

In the color-confining 
light-front holographic model discussed here, the vanishing of the pion mass in the chiral limit, a phenomenon usually ascribed to spontaneous symmetry breaking of the chiral symmetry,  is  obtained specifically from the precise cancellation of the LF kinetic energy and LF potential energy terms for the quadratic confinement potential. This mechanism provides a  viable alternative to the conventional description of nonperturbative QCD based on vacuum condensates, and it 
eliminates a major conflict of hadron physics with the empirical value for the cosmological constant~\cite{Brodsky:2009zd,  Brodsky:2010xf}.

\section{Acknowledgments}
Invited talk, presented by SJB at the	
{\it International Conference on Nuclear Theory in the Supercomputing Era } (NTSE 2013), 	
 in honor of James Vary,  May 13 - May 17, 2013,   Iowa State University, Ames, Iowa.
This work  was supported by the Department of Energy contract DE--AC02--76SF00515.   
SLAC-PUB-15735


\begin{thebibliography}{0}


\bibitem{Dirac:1949cp} 
  P.~A.~M.~Dirac,
  Rev.\ Mod.\ Phys.\  {\bf 21}, 392 (1949).
  
  
\bibitem{Lorce:2012jy} 
  C.~Lorce and B.~Pasquini,
  Int.\ J.\ Mod.\ Phys.\ Conf.\ Ser.\  {\bf 20}, 84 (2012)
  [arXiv:1208.3065 [hep-ph]].
  
  
\bibitem{Brodsky:2000xy} 
  S.~J.~Brodsky, M.~Diehl and D.~S.~Hwang,
  Nucl.\ Phys.\ B {\bf 596}, 99 (2001)
  [hep-ph/0009254].


\bibitem{Drell:1969km} 
  S.~D.~Drell and T.~-M.~Yan,
  Phys.\ Rev.\ Lett.\  {\bf 24}, 181 (1970).


\bibitem{West:1970av} 
  G.~B.~West,
  Phys.\ Rev.\ Lett.\  {\bf 24}, 1206 (1970).


\bibitem{Brodsky:1980zm} 
  S.~J.~Brodsky and S.~D.~Drell,
  Phys.\ Rev.\ D {\bf 22}, 2236 (1980).


\bibitem{Brodsky:1968ea} 
  S.~J.~Brodsky and J.~R.~Primack,
  Annals Phys.\  {\bf 52}, 315 (1969).


\bibitem{Brodsky:1968xc} 
  S.~J.~Brodsky and J.~R.~Primack,
  Phys.\ Rev.\  {\bf 174}, 2071 (1968).
  
\bibitem{Lepage:1980fj} 
  G.~P.~Lepage and S.~J.~Brodsky,
  Phys.\ Rev.\ D {\bf 22}, 2157 (1980).


\bibitem{Efremov:1979qk} 
  A.~V.~Efremov and A.~V.~Radyushkin,
  Phys.\ Lett.\ B {\bf 94}, 245 (1980).



\bibitem{Brodsky:1973kb} 
  S.~J.~Brodsky, R.~Roskies and R.~Suaya,
  Phys.\ Rev.\ D {\bf 8}, 4574 (1973).


\bibitem{Feynman:1969ej} 
  R.~P.~Feynman,
  Phys.\ Rev.\ Lett.\  {\bf 23}, 1415 (1969).


\bibitem{Feynman:1973xc} 
  R.~P.~Feynman,
  Reading 1972, 282p.
  
 
  \bibitem{Fubini:1965xx}
 S.~Fubini and G.~Furlan,
 ``Renormalization effects for partially conserved currents,'' 
 Physics {\bf 1}, 229 (1965). 


\bibitem{Weinberg:1966jm} 
  S.~Weinberg,
  Phys.\ Rev.\  {\bf 150}, 1313 (1966).


\bibitem{Mueller:1993rr} 
  A.~H.~Mueller,
  Nucl.\ Phys.\ B {\bf 415}, 373 (1994).


\bibitem{Brodsky:1985gs} 
  S.~J.~Brodsky and C.~-R.~Ji,
  Phys.\ Rev.\ D {\bf 33}, 2653 (1986).


\bibitem{Brodsky:2000ii} 
  S.~J.~Brodsky, D.~S.~Hwang, B.~-Q.~Ma and I.~Schmidt,
  Nucl.\ Phys.\ B {\bf 593}, 311 (2001)
  [hep-th/0003082].


\bibitem{Cruz-Santiago:2013vta} 
  C.~A.~Cruz-Santiago and A.~M.~Stasto,
  Nucl.\ Phys.\ B {\bf 875}, 368 (2013)
  [arXiv:1308.1062 [hep-ph]].


\bibitem{Antonuccio:1997tw} 
  F.~Antonuccio, S.~J.~Brodsky and S.~Dalley,
  Phys.\ Lett.\ B {\bf 412}, 104 (1997)
  [hep-ph/9705413].


\bibitem{Brodsky:1994kg} 
  S.~J.~Brodsky, M.~Burkardt and I.~Schmidt,
  Nucl.\ Phys.\ B {\bf 441}, 197 (1995)
  [hep-ph/9401328].


\bibitem{Gutsche:2013zia} 
  T.~Gutsche, V.~E.~Lyubovitskij, I.~Schmidt and A.~Vega,
  arXiv:1306.0366 [hep-ph].


\bibitem{Brodsky:2002ue} 
  S.~J.~Brodsky, P.~Hoyer, N.~Marchal, S.~Peigne and F.~Sannino,
  Phys.\ Rev.\ D {\bf 65}, 114025 (2002)
  [hep-ph/0104291].


\bibitem{Brodsky:2002cx} 
  S.~J.~Brodsky, D.~S.~Hwang and I.~Schmidt,
  Phys.\ Lett.\ B {\bf 530}, 99 (2002)
  [hep-ph/0201296].
  

\bibitem{Brodsky:1988xz} S.~J.~Brodsky and A.~H.~Mueller,
  Phys.\ Lett.\ B {\bf 206}, 685 (1988).
  


\bibitem{Brodsky:1983vf} 
  S.~J.~Brodsky, C.~-R.~Ji and G.~P.~Lepage,
  Phys.\ Rev.\ Lett.\  {\bf 51}, 83 (1983).


\bibitem{Brodsky:1980pb} 
  S.~J.~Brodsky, P.~Hoyer, C.~Peterson and N.~Sakai,
  Phys.\ Lett.\ B {\bf 93}, 451 (1980).


\bibitem{Franz:2000ee} 
  M.~Franz, M.~V.~Polyakov and K.~Goeke,
  Phys.\ Rev.\ D {\bf 62}, 074024 (2000)
  [hep-ph/0002240].


\bibitem{Brodsky:2008tk} 
  S.~J.~Brodsky, G.~F.~de Teramond and R.~Shrock,
  AIP Conf.\ Proc.\  {\bf 1056}, 3 (2008)
  [arXiv:0807.2484 [hep-ph]].


\bibitem{Brodsky:1997de} 
  S.~J.~Brodsky, H.~-C.~Pauli and S.~S.~Pinsky,
  Phys.\ Rept.\  {\bf 301}, 299 (1998)
  [hep-ph/9705477].


\bibitem{Bardeen:1976tm} 
  W.~A.~Bardeen and R.~B.~Pearson,
  Phys.\ Rev.\ D {\bf 14}, 547 (1976).
  
  
\bibitem{Burkardt:2001mf} 
  M.~Burkardt and S.~K.~Seal,
  Phys.\ Rev.\ D {\bf 65}, 034501 (2002)
  [hep-ph/0102245].
  
\bibitem{Bratt:2004wq} 
  J.~Bratt, S.~Dalley, B.~van de Sande and E.~M.~Watson,
  Phys.\ Rev.\ D {\bf 70}, 114502 (2004)
  [hep-ph/0410188].



\bibitem{Pauli:1985pv} 
  H.~C.~Pauli and S.~J.~Brodsky,
  Phys.\ Rev.\ D {\bf 32}, 1993 (1985).


\bibitem{Hellerman:1999nr} 
  S.~Hellerman and J.~Polchinski,
  In *Shifman, M.A. (ed.): The many faces of the superworld* 142-155
  [hep-th/9908202].


\bibitem{Polchinski:1999br} 
  J.~Polchinski,
  Prog.\ Theor.\ Phys.\ Suppl.\  {\bf 134}, 158 (1999)
  [hep-th/9903165].


\bibitem{Vary:2009gt} 
  J.~P.~Vary, H.~Honkanen, J.~Li, P.~Maris, S.~J.~Brodsky, A.~Harindranath, G.~F.~de Teramond and P.~Sternberg {\it et al.},
  Phys.\ Rev.\ C {\bf 81}, 035205 (2010)
  [arXiv:0905.1411 [nucl-th]].


\bibitem{Parisi:1972zy} 
  G.~Parisi,
  Phys.\ Lett.\ B {\bf 39}, 643 (1972).


 \bibitem{deAlfaro:1976je}
  V.~de Alfaro, S.~Fubini and G.~Furlan,
Nuovo Cim.\ A {\bf 34}, 569 (1976).



  \bibitem{Brodsky:2013kpr} 
  S.~J.~Brodsky, G.~F.~de Teramond and H.~G.~Dosch,
  arXiv:1302.4105 [hep-th].
  
  
\bibitem{Glazek:2013jba} 
  S.~D.~Glazek and A.~P.~Trawinski,
  arXiv:1307.2059 [hep-ph].


\bibitem{Karch:2006pv} 
  A.~Karch, E.~Katz, D.~T.~Son and M.~A.~Stephanov,
  Phys.\ Rev.\ D {\bf 74}, 015005 (2006)
  [hep-ph/0602229].


\bibitem{Polchinski:2001tt} 
  J.~Polchinski and M.~J.~Strassler,
  Phys.\ Rev.\ Lett.\  {\bf 88}, 031601 (2002)
  [hep-th/0109174].


\bibitem{Brodsky:1973kr} 
  S.~J.~Brodsky and G.~R.~Farrar,
  Phys.\ Rev.\ Lett.\  {\bf 31}, 1153 (1973).
  
\bibitem{Matveev:1973ra} 
  V.~A.~Matveev, R.~M.~Muradian and A.~N.~Tavkhelidze,
  Lett.\ Nuovo Cim.\  {\bf 7}, 719 (1973).


\bibitem{deTeramond:2008ht} 
  G.~F.~de Teramond and S.~J.~Brodsky,
  Phys.\ Rev.\ Lett.\  {\bf 102}, 081601 (2009)
  [arXiv:0809.4899 [hep-ph]].


\bibitem{Maldacena:1997re} 
  J.~M.~Maldacena,
  Adv.\ Theor.\ Math.\ Phys.\  {\bf 2}, 231 (1998)
  [hep-th/9711200].


\bibitem{Polchinski:2002jw} 
  J.~Polchinski and M.~J.~Strassler,
  JHEP {\bf 0305}, 012 (2003)
  [hep-th/0209211].


\bibitem{Abidin:2008ku} 
  Z.~Abidin and C.~E.~Carlson,
  Phys.\ Rev.\ D {\bf 77}, 095007 (2008)
  [arXiv:0801.3839 [hep-ph]].


\bibitem{Brodsky:2006uqa} 
  S.~J.~Brodsky and G.~F.~de Teramond,
  Phys.\ Rev.\ Lett.\  {\bf 96}, 201601 (2006)
  [hep-ph/0602252].


\bibitem{Brodsky:2007hb} 
  S.~J.~Brodsky and G.~F.~de Teramond,
  Phys.\ Rev.\ D {\bf 77}, 056007 (2008)
  [arXiv:0707.3859 [hep-ph]].


\bibitem{Brodsky:2008pf} 
  S.~J.~Brodsky and G.~F.~de Teramond,
  Phys.\ Rev.\ D {\bf 78}, 025032 (2008)
  [arXiv:0804.0452 [hep-ph]].


\bibitem{deTeramond:2013it} 
  G.~F.~de Teramond, H.~G.~Dosch and S.~J.~Brodsky,
  Phys.\ Rev.\ D {\bf 87}, 075005 (2013)
  [arXiv:1301.1651 [hep-ph]].


\bibitem{deTeramond:2009xk} 
  G.~F.~de Teramond and S.~J.~Brodsky,
  Nucl.\ Phys.\ Proc.\ Suppl.\  {\bf 199}, 89 (2010)
  [arXiv:0909.3900 [hep-ph]].


\bibitem{Soper:1976jc} 
  D.~E.~Soper,
  Phys.\ Rev.\ D {\bf 15}, 1141 (1977).


\bibitem{Pauli:1998tf} 
  H.~C.~Pauli,
  Eur.\ Phys.\ J.\ C {\bf 7}, 289 (1999)
  [hep-th/9809005].


\bibitem{Chabysheva:2011ed} 
  S.~S.~Chabysheva and J.~R.~Hiller,
  Phys.\ Lett.\ B {\bf 711}, 417 (2012)
  [arXiv:1103.0037 [hep-ph]].


\bibitem{deTeramond:2012rt} 
  G.~F.~de Teramond and S.~J.~Brodsky,
  arXiv:1203.4025 [hep-ph].


\bibitem{Gutsche:2011vb} 
  T.~Gutsche, V.~E.~Lyubovitskij, I.~Schmidt and A.~Vega,
  Phys.\ Rev.\ D {\bf 85}, 076003 (2012)
  [arXiv:1108.0346 [hep-ph]].


\bibitem{deTeramond:2010ge} 
  G.~F.~de Teramond and S.~J.~Brodsky,
  AIP Conf.\ Proc.\  {\bf 1296}, 128 (2010)
  [arXiv:1006.2431 [hep-ph]].


\bibitem{Breitenlohner:1982jf} 
  P.~Breitenlohner and D.~Z.~Freedman,
  Annals Phys.\  {\bf 144}, 249 (1982).


\bibitem{Appelquist:1977tw} 
  T.~Appelquist, M.~Dine and I.~J.~Muzinich,
  Phys.\ Lett.\ B {\bf 69}, 231 (1977).
  
  
\bibitem{Bjorken:2013boa} 
  J.~D.~Bjorken, S.~J.~Brodsky and A.~S.~Goldhaber,
  arXiv:1308.1435 [hep-ph].


\bibitem{Srivastava:2002mw} 
  P.~P.~Srivastava and S.~J.~Brodsky,
  Phys.\ Rev.\ D {\bf 66}, 045019 (2002)
  [hep-ph/0202141].


\bibitem{Casher:1974xd} 
  A.~Casher and L.~Susskind,
  Phys.\ Rev.\ D {\bf 9}, 436 (1974).


\bibitem{Brodsky:2009zd} 
  S.~J.~Brodsky and R.~Shrock,
  Proc.\ Nat.\ Acad.\ Sci.\  {\bf 108}, 45 (2011)
  [arXiv:0905.1151 [hep-th]].


\bibitem{Brodsky:2010xf} 
  S.~J.~Brodsky, C.~D.~Roberts, R.~Shrock and P.~C.~Tandy,
  Phys.\ Rev.\ C {\bf 82}, 022201 (2010)
  [arXiv:1005.4610 [nucl-th]].


\bibitem{Chang:2013pq} 
  L.~Chang, I.~C.~Cloet, J.~J.~Cobos-Martinez, C.~D.~Roberts, S.~M.~Schmidt and P.~C.~Tandy,
  arXiv:1301.0324 [nucl-th].
  
\bibitem{Chang:2013epa} 
  L.~Chang, C.~D.~Roberts and S.~M.~Schmidt,
  arXiv:1308.4708 [nucl-th].
  
\bibitem{Glazek:2011vg} 
  S.~D.~Glazek,
  Acta Phys.\ Polon.\ B {\bf 42}, 1933 (2011)
  [arXiv:1106.6100 [hep-th]].




\bibitem{Brodsky:2011ig} 
  S.~J.~Brodsky and L.~Di Giustino,
  Phys.\ Rev.\ D {\bf 86}, 085026 (2012)
  [arXiv:1107.0338 [hep-ph]].


\bibitem{Brodsky:1982gc} 
  S.~J.~Brodsky, G.~P.~Lepage and P.~B.~Mackenzie,
  Phys.\ Rev.\ D {\bf 28}, 228 (1983).


\bibitem{Mojaza:2012mf} 
  M.~Mojaza, S.~J.~Brodsky and X.~-G.~Wu,
  Phys.\ Rev.\ Lett.\  {\bf 110}, 192001 (2013)
  [arXiv:1212.0049 [hep-ph]].


\bibitem{Brodsky:2013vpa} 
  S.~J.~Brodsky, M.~Mojaza and X.~-G.~Wu,
  arXiv:1304.4631 [hep-ph].


\bibitem{Brodsky:2012ik} 
  S.~J.~Brodsky and X.~-G.~Wu,
  Phys.\ Rev.\ D {\bf 85}, 114040 (2012)
  [arXiv:1205.1232 [hep-ph]].


\bibitem{Brodsky:1994eh} 
  S.~J.~Brodsky and H.~J.~Lu,
  Phys.\ Rev.\ D {\bf 51}, 3652 (1995)
  [hep-ph/9405218].


\bibitem{Brodsky:1995tb} 
  S.~J.~Brodsky, G.~T.~Gabadadze, A.~L.~Kataev and H.~J.~Lu,
  Phys.\ Lett.\ B {\bf 372}, 133 (1996)
  [hep-ph/9512367].


\bibitem{Wu:2013ei} 
  X.~-G.~Wu, S.~J.~Brodsky and M.~Mojaza,
  Prog.\ Part.\ Nucl.\ Phys.\  {\bf 72}, 44 (2013)
  [arXiv:1302.0599 [hep-ph]].


\bibitem{Weinberg:1967kj} 
  S.~Weinberg,
  Phys.\ Rev.\ Lett.\  {\bf 18}, 507 (1967).


\bibitem{Brodsky:2010ur} 
  S.~J.~Brodsky, G.~F.~de Teramond and A.~Deur,
  Phys.\ Rev.\ D {\bf 81}, 096010 (2010)
  [arXiv:1002.3948 [hep-ph]].


\bibitem{Forshaw:2012im} 
  J.~R.~Forshaw and R.~Sandapen,
  Phys.\ Rev.\ Lett.\  {\bf 109}, 081601 (2012)
  [arXiv:1203.6088 [hep-ph]].


\bibitem{Vega:2010ns} 
  A.~Vega, I.~Schmidt, T.~Gutsche and V.~E.~Lyubovitskij,
  Phys.\ Rev.\ D {\bf 83}, 036001 (2011)
  [arXiv:1010.2815 [hep-ph]].


\bibitem{Chakrabarti:2013dda} 
  D.~Chakrabarti and C.~Mondal,
  arXiv:1307.7995 [hep-ph].


\bibitem{Shifman:1978bx} 
  M.~A.~Shifman, A.~I.~Vainshtein and V.~I.~Zakharov,
  Nucl.\ Phys.\ B {\bf 147}, 385 (1979).


\bibitem{Maris:1997hd} 
  P.~Maris, C.~D.~Roberts and P.~C.~Tandy,
  Phys.\ Lett.\ B {\bf 420}, 267 (1998)
  [nucl-th/9707003].


\bibitem{Brodsky:2012ku} 
  S.~J.~Brodsky, C.~D.~Roberts, R.~Shrock and P.~C.~Tandy,
  Phys.\ Rev.\ C {\bf 85}, 065202 (2012)
  [arXiv:1202.2376 [nucl-th]].


\bibitem{GellMann:1968rz} 
  M.~Gell-Mann, R.~J.~Oakes and B.~Renner,
  Phys.\ Rev.\  {\bf 175}, 2195 (1968).


\end{thebibliography}
\end{document}